\documentclass[reqno,11pt]{amsart}
\usepackage{amscd}
\usepackage{slashed}
\usepackage{amssymb}
\usepackage{ushort}
\usepackage{mathtools}
\usepackage{esint}
\usepackage{enumitem}
\usepackage{pstricks}
\usepackage{pst-grad} 
\usepackage{pst-plot} 
\usepackage[mathscr]{eucal}
\numberwithin{equation}{section}
\allowdisplaybreaks[1]

\title[Lectures on Linear Stability of Rotating Black Holes]{Lectures on Linear Stability of \\
Rotating Black Holes}

\author[F.\ Finster]{Felix Finster \\ \\ November 2018}
\address{Fakult\"at f\"ur Mathematik \\ Universit\"at Regensburg \\ D-93040 Regensburg \\ Germany}
\email{finster@ur.de}

\newtheorem{Def}{Def.}[section]
\newtheorem{Thm}[Def]{Theorem}
\newtheorem{Prp}[Def]{Proposition}
\newtheorem{Lemma}[Def]{Lemma}

\newcommand{\Thanks}{\vspace*{.5em} \noindent \thanks}
\newcommand{\Proof}{\begin{proof}}
\newcommand{\QED}{\end{proof} \noindent}

\newcommand{\C}{\mathbb{C}}
\newcommand{\R}{\mathbb{R}}
\newcommand{\1}{\mbox{\rm 1 \hspace{-1.05 em} 1}}
\newcommand{\Z}{\mathbb{Z}}

\newcommand{\N}{\mathbb{N}}

\newcommand{\beq}{\begin{equation}}
\newcommand{\eeq}{\end{equation}}

\newcommand{\A}{\mathcal{A}}

\newcommand{\phiD}{\phi^\text{\tiny{\rm{D}}}}

\renewcommand{\H}{\mathscr{H}}
\newcommand{\la}{\langle}
\newcommand{\ra}{\rangle}
\newcommand{\Lin}{\text{\rm{L}}}

\newcommand{\D}{\mathscr{D}}
\newcommand{\scrM}{\mycal M}
\newcommand{\scrN}{\mycal N}
\newcommand{\Sact}{{\mathcal{S}}}
\renewcommand{\L}{{\mathcal{L}}}
\newcommand\B{{\mathscr{B}}}
\newcommand{\Sl}{\mbox{$\prec \!\!$ \nolinebreak}}
\newcommand{\Sr}{\mbox{\nolinebreak $\succ$}}
\newcommand{\bitem}{\begin{itemize}[leftmargin=2em]}
\newcommand{\eitem}{\end{itemize}}
\newcommand{\itemD}{\item[{\raisebox{0.125em}{\tiny $\blacktriangleright$}}]}

\DeclareFontFamily{OT1}{rsfso}{}
\DeclareFontShape{OT1}{rsfso}{m}{n}{ <-7> rsfso5 <7-10> rsfso7 <10-> rsfso10}{}
\DeclareMathAlphabet{\mycal}{OT1}{rsfso}{m}{n}

\DeclareMathOperator{\im}{Im}

\begin{document}
\maketitle


\begin{abstract}
These lecture notes are concerned with linear stability of
the non-extreme Kerr geometry under perturbations of general spin.
After a brief review of the Kerr black hole
and its symmetries, we describe these symmetries by Killing fields
and work out the connection to conservation laws.
The Penrose process and superradiance effects are discussed.
Decay results on the long-time behavior of Dirac waves are outlined.
It is explained schematically how the Maxwell equations and the equations for
linearized gravitational waves can be decoupled to obtain the Teukolsky equation.
It is shown how the Teukolsky equation can be fully separated to
a system of coupled ordinary differential equations.
Linear stability of the non-extreme Kerr black hole is stated as a
pointwise decay result for solutions of the Cauchy problem
for the Teukolsky equation. The stability proof is outlined,
with an emphasis on the underlying ideas and methods.
\end{abstract}

\tableofcontents

\newpage
\section{Introduction}
These lectures are concerned with the black hole stability problem.
Since this is a broad topic which many people have been working on,
we shall restrict attention to specific aspects of this problem:
First, we will be concerned only with {\em{linear}} stability.
Indeed, the problem of nonlinear stability is much harder, and at present
there are only few rigorous results. Second, we will concentrate on {\em{rotating}}
black holes. This is because the angular momentum leads to effects
(Penrose process, superradiance) which make the rotating case particularly interesting.
Moreover, the focus on rotating black holes gives a better connection to my own research,
which was carried out in collaboration with Niky Kamran (McGill), Joel Smoller (University of
Michigan) and Shing-Tung Yau (Harvard). The linear stability result for general spin
was obtained together with Joel Smoller (see~\cite{stable} and the survey article~\cite{cmsa}).
Before beginning, I would like to remember Joel Smoller, who sadly passed away in September 2017.
These notes are dedicated to his memory.

\section{The Kerr Black Hole}
In general relativity, space and time are combined to a four-dimensional space-time,
which is modelled mathematically by a Lorentzian manifold~$(\scrM, g)$ of signature $(+ \ \!\! - \ \!\! - \ \! - )$
(for more elementary or more detailed introductions to general relativity
see the textbooks~\cite{adler, misner, wald, straumann2}).
The gravitational field is described geometrically in terms of
the curvature of space-time. Newton's gravitational law is replaced by the Einstein equations
\beq \label{einstein}
R_{jk} - \frac{1}{2}\: R\, g_{jk} = 8 \pi \kappa\, T_{jk} \:,
\eeq
where~$R_{jk}$ is the Ricci tensor, $R$ is scalar curvature, and~$\kappa$ denotes the gravitational constant.
Here~$T_{jk}$ is the energy-momentum tensor which describes the distribution of matter in space-time.

A rotating black hole is described by the {\em{Kerr geometry}}. It is
a solution of the vacuum Einstein equations discovered in 1963 by Roy Kerr.
In the so-called Boyer-Lindquist coordinates, the Kerr metric takes the form
(see~\cite{chandra, oneill})
\beq \label{BL}
ds^{2} = \frac{\Delta}{U} \,( dt-a\sin^{2}\vartheta\,d\varphi)^{2}-U \left( \frac{dr^{2}}{\Delta}+d\vartheta^{2}
\right) -\frac{\sin^{2}\vartheta}{U} \Big( a\,dt-(r^{2}+a^{2})d\varphi \Big)^{2} \:,
\eeq
where
\beq \label{UDdef}
U = r^{2}+a^{2}\cos^{2}\vartheta,\qquad \Delta = r^{2}-2Mr+a^{2} \:,
\eeq
and the coordinates~$(t,r, \vartheta, \varphi)$ are in the range
\[ -\infty<t<\infty,\;\;\;
M+{\sqrt{M^{2}-a^{2}}}<r<\infty,\;\;\; 0<\vartheta<\pi,\;\;\;
0<\varphi<2\pi \:. \]
The parameters~$M$ and~$aM$ describe the mass and the angular
momentum of the black hole.

In the case~$a=0$, one recovers the Schwarzschild metric
\[ ds^{2} = \left(1-\frac{2M}{r} \right) dt^{2}-
\left(1-\frac{2M}{r} \right)^{-1} dr^2 -r^{2}(d\theta^{2}+\sin^{2}\theta\,d\varphi^{2})\:. \]
In this case, the function~$\Delta$ has two roots
\begin{align*}
r&=2M &&\hspace*{-1cm} \text{event horizon} \\
r&=0 &&\hspace*{-1cm} \text{curvature singularity} \:.
\end{align*}
In the region~$r>2M$, the so-called {\em{exterior region}}, $t$ is a time coordinate,
whereas~$r$, $\vartheta$ and~$\varphi$ are spatial coordinates.
More precisely, $(\vartheta, \varphi)$ are polar coordinates, whereas the radial coordinate~$r$
is determined by the fact that the two-surface $S=\{t=t_0, r=r_0\}$ has area~$4 \pi r_0^2$.
The region~$r<2M$, on the other hand, is the {\em{interior region}}.
In this region, the radial coordinate~$r$ is time, whereas~$t$ is a spatial coordinate.
Since time always propagates to the future, the event horizon can be regarded as the ``boundary of no escape.''
The surface~$r=2M$ merely is a coordinate singularity of our metric.
This becomes apparent by transforming to Eddington-Finkelstein or Kruskal coordinates.
For brevity, we shall not enter the details here.

In the case~$a \neq 0$, the singularity structure is more involved.
The function~$U$ is always strictly positive.
The function~$\Delta$ has the two roots
\begin{align}
r_0&:= M + \sqrt{M^2-a^2} &&\hspace*{-1cm} \text{event horizon} \label{EH} \\
r_1&:= M - \sqrt{M^2-a^2} &&\hspace*{-1cm} \text{Cauchy horizon} \:.
\end{align}
If~$a^2>M^2$, these roots are complex. This corresponds to the unphysical situation
of a naked singularity. We shall not discuss this case here, but only consider the so-called
\[ \text{\em{non-extreme case}} \qquad M^2 < a^2 \:. \]
In this case, the hypersurface
\[ r = r_{1} := M+{\sqrt{M^{2}-a^{2}}} \]
again defines the {\em{event horizon}} of the black hole.
In what follows, we shall restrict attention to the {\em{exterior region}}~$r>r_1$.
This is because classically, no information can be transmitted from the interior of the
black hole to its exterior. Therefore, it is impossible for principal reasons to know what
happens inside the black hole. With this in mind, it seems pointless to study the
black hole inside the event horizon, because this study will never be tested or verified
by experiments.

We finally remark that in {\em{quantum gravity}}, the situation is quite different because
it is conceivable that a black hole might ``evaporate,'' in which case the interior of the black
hole might become accessible to observations.
In physics, such questions are often discussed in connection with the so-called information paradox,
which states that the loss of information at the event horizon is not compatible with the unitary
time evolution in quantum theory.
I find such questions related to quantum effects of a black hole quite interesting, 
and indeed most of my recent research is devoted to quantum gravity
(in an approach called causal fermion systems; see for example the textbook~\cite{cfs}
or the survey paper~\cite{dice2014}).
But since this summer school is devoted to classical gravity, I shall not enter this topic here.

\section{Symmetries and Killing Fields}
The Kerr geometry is stationary and axisymmetric. This is apparent in Boyer-Lindquist coordinates~\eqref{BL}
because the metric coefficients are
\begin{align*}
&\text{independent of~$t$:} && \hspace*{-3cm} \text{stationary} \\
&\text{independent of~$\varphi$:} && \hspace*{-3cm} \text{axisymmetric}\:.
\end{align*}
These symmetries can be described more abstractly using the notion of
{\em{Killing fields}}. We recall how this works because we need it later for the description
of the Penrose process and superradiance.
We restrict attention to the time translation symmetry, because for for the axisymmetry or other symmetries,
the argument is similar. Given~$\tau \in \R$, we consider the mapping
\[ \Phi_\tau \::\: \scrM \rightarrow \scrM \:,\qquad (t,x) \mapsto (t+\tau, x) \]
(where~$x$ stands for the spatial coordinates~$(r, \vartheta, \varphi)$).
The fact that the metric coefficients are time independent means that~$\Phi_\tau$ is an {\em{isometry}},
defined as follows. The derivative of~$\Phi_\tau$ (i.e.\ the linearization; it is sometimes
also denoted by~$(\Phi_\tau)_*$) is a mapping between the corresponding tangent spaces,
\[ 
D\Phi_\tau|_p \::\: T_p\scrM \rightarrow T_{\Phi_\tau(x)}\scrM \:. \]
Being an isometry means that
\[ g_p(u,v) = g_{\Phi_\tau(p)}\big( D\Phi_\tau|_p u, D\Phi_\tau|_p v \big) \qquad
\text{ for all~$u,v \in T_p\scrM$}\:. \]
Let us evaluate this equation infinitesimally in~$\tau$. We first introduce the vector field~$K$ by
\[ K := \frac{d}{d\tau} \Phi_\tau \big|_{\tau=0} \:. \]
Choosing local coordinates, we obtain in components
\[ \big( D\Phi_\tau|_p u \big)^a = \frac{\partial \Phi^a_\tau(p)}{\partial x^i}\: u^i \:, \]
where for clarity we denote the tensor indices at the point~$\Phi_\tau(x)$ by~$a$ and~$b$. We then obtain
\begin{align*}
0 &= \frac{d}{d\tau} \:g_{\Phi_\tau(p)}\big( D\Phi_\tau|_p u, D\Phi_\tau|_p v \big) \Big|_{\tau=0} \\
&= \frac{d}{d\tau} \:\Big( g_{ab}\big(\Phi_\tau(p)\big)\: 
\frac{\partial \Phi^a_\tau(p)}{\partial x^i}\: u^i \: \frac{\partial \Phi^b_\tau(p)}{\partial x^j}\: v^j
\Big) \Big|_{\tau=0} \\
&= \partial_k g(u,v)\: K^k + g\big( u^i \partial_i K, v \big) + g\big( u, v^j \partial_j K \big) \:.
\end{align*}
Choosing Gaussian coordinates at~$p$, one sees that this equation can be written covariantly as
\[ 0 = g\big( \nabla_u K, v \big) + g\big( u, \nabla_v  K \big) \:, \]
where~$\nabla$ is the Levi-Civita connection.
This is the {\em{Killing equation}}, which can also be written in the shorter form
\beq \label{killing}
0 = \nabla_{(i} K_{j)} := \frac{1}{2}\:\big( \nabla_i K_j + \nabla_j K_i \big) \:.
\eeq
A vector field which satisfies the Killing equation is referred to as a {\em{Killing field}}.
We remark that if the flow lines exist on an interval containing zero and~$\tau$, then the
resulting diffeomorphism~$\Phi_\tau$ is indeed an isometry of~$\scrM$.

A variant of Noether's theorem states that Killing symmetries, which describe infinitesimal
symmetries of space-time, give rise to corresponding conservation laws.
For {\em{geodesics}}, these conservation laws are obtained simply by taking the Lorentzian inner product
of the Killing vector field and the velocity vector of the geodesic. Indeed, let~$\gamma(\tau)$
be a parametrized geodesic, i.e.
\[ \nabla_\tau \dot{\gamma}(\tau) = 0 \:. \]
Then, denoting the metric for simplicity by~$\la .,. \ra_p := g_p(.,.)$, we obtain
\begin{align*}
\frac{d}{d\tau} &\big\la K(\gamma(\tau)), \dot{\gamma}(\tau) \big\ra_{\gamma(\tau)} 
= \big\la \nabla_\tau K(\gamma(\tau)), \dot{\gamma}(\tau) \big\ra_{\gamma(\tau)} 
+ \big\la K(\gamma(\tau)), \nabla_\tau \dot{\gamma}(\tau) \big\ra_{\gamma(\tau)} \\
&= \big\la \nabla_\tau K(\gamma(\tau)), \dot{\gamma}(\tau) \big\ra_{\gamma(\tau)} 
= \nabla_i K_j \big|_{\gamma(\tau)}\: \dot{\gamma}^i(\tau)\: \dot{\gamma}^j(\tau) = 0 \:,
\end{align*}
where in the last step we used the Killing equation~\eqref{killing}.
We thus obtain the {\em{conservation law}}
\[ 
\big\la K(\gamma(\tau)), \dot{\gamma}(\tau) \big\ra_{\gamma(\tau)} = \text{const} \:,
\] 
which holds for any parametrized geodesic~$\gamma(\tau)$ and any Killing field~$K$.

\section{The Penrose Process and Superradiance}
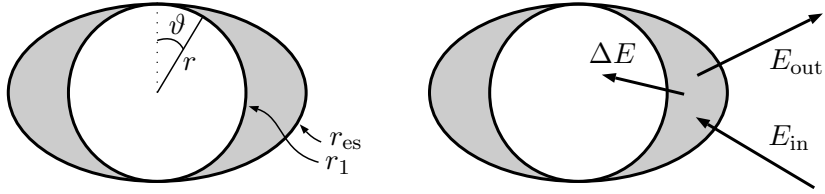
\begin{figure}
%
\psscalebox{1.0 1.0} 
{
\begin{pspicture}(-1.8,-1.2410852)(14.975,1.2410852)
\definecolor{colour0}{rgb}{0.8,0.8,0.8}
\pspolygon[linecolor=colour0, linewidth=0.02, fillstyle=solid,fillcolor=colour0](7.905,1.1960852)(8.115,1.1810852)(8.425,1.1110852)(8.825,0.9610852)(9.255,0.6710852)(9.435,0.47608522)(9.545,0.2210852)(9.575,-0.008914795)(9.505,-0.2589148)(9.355,-0.50891477)(9.13,-0.69891477)(8.815,-0.88391477)(8.56,-0.9839148)(8.24,-1.0689148)(7.925,-1.1089149)(8.17,-1.0139148)(8.44,-0.8089148)(8.62,-0.57391477)(8.725,-0.3439148)(8.78,-0.1189148)(8.785,0.1460852)(8.725,0.4460852)(8.535,0.7860852)(8.305,1.0110852)(8.07,1.1360852)
\pspolygon[linecolor=colour0, linewidth=0.02, fillstyle=solid,fillcolor=colour0](7.295,-1.1189148)(7.085,-1.1039147)(6.775,-1.0339148)(6.375,-0.88391477)(5.945,-0.5939148)(5.765,-0.39891478)(5.655,-0.14391479)(5.625,0.08608521)(5.695,0.3360852)(5.845,0.5860852)(6.07,0.7760852)(6.385,0.9610852)(6.64,1.0610852)(6.96,1.1460853)(7.275,1.1860852)(7.03,1.0910852)(6.76,0.8860852)(6.58,0.6510852)(6.475,0.4210852)(6.42,0.1960852)(6.415,-0.06891479)(6.475,-0.36891478)(6.665,-0.7089148)(6.895,-0.9339148)(7.13,-1.0589148)
\pspolygon[linecolor=colour0, linewidth=0.02, fillstyle=solid,fillcolor=colour0](2.305,1.1960852)(2.515,1.1810852)(2.825,1.1110852)(3.225,0.9610852)(3.655,0.6710852)(3.835,0.47608522)(3.945,0.2210852)(3.975,-0.008914795)(3.905,-0.2589148)(3.755,-0.50891477)(3.53,-0.69891477)(3.215,-0.88391477)(2.96,-0.9839148)(2.64,-1.0689148)(2.325,-1.1089149)(2.57,-1.0139148)(2.84,-0.8089148)(3.02,-0.57391477)(3.125,-0.3439148)(3.18,-0.1189148)(3.185,0.1460852)(3.125,0.4460852)(2.935,0.7860852)(2.705,1.0110852)(2.47,1.1360852)
\pspolygon[linecolor=colour0, linewidth=0.02, fillstyle=solid,fillcolor=colour0](1.695,-1.1189148)(1.485,-1.1039147)(1.175,-1.0339148)(0.775,-0.88391477)(0.345,-0.5939148)(0.165,-0.39891478)(0.055,-0.14391479)(0.025,0.08608521)(0.095,0.3360852)(0.245,0.5860852)(0.47,0.7760852)(0.785,0.9610852)(1.04,1.0610852)(1.36,1.1460853)(1.675,1.1860852)(1.43,1.0910852)(1.16,0.8860852)(0.98,0.6510852)(0.875,0.4210852)(0.82,0.1960852)(0.815,-0.06891479)(0.875,-0.36891478)(1.065,-0.7089148)(1.295,-0.9339148)(1.53,-1.0589148)
\pscircle[linecolor=black, linewidth=0.04, dimen=outer](2.0,0.041085206){1.2}
\psellipse[linecolor=black, linewidth=0.04, dimen=outer](2.0,0.041085206)(2.0,1.2)
\psline[linecolor=black, linewidth=0.02](2.0,0.041085206)(2.6,1.0410852)
\psline[linecolor=black, linewidth=0.02, linestyle=dotted, dotsep=0.10583334cm](2.0,0.041085206)(2.0,1.2410852)
\psbezier[linecolor=black, linewidth=0.02](2.0,0.7360852)(2.15,0.7860852)(2.28,0.7560852)(2.35,0.616085205078125)
\rput[bl](2.155,0.8260852){\normalsize{$\vartheta$}}
\rput[bl](2.34,0.3410852){\normalsize{$r$}}
\psbezier[linecolor=black, linewidth=0.02, arrowsize=0.05291667cm 2.0,arrowlength=1.4,arrowinset=0.0]{->}(4.145,-0.88391477)(3.63,-0.7439148)(3.68,-0.2889148)(3.215,-0.138914794921875)
\psbezier[linecolor=black, linewidth=0.02, arrowsize=0.05291667cm 2.0,arrowlength=1.4,arrowinset=0.0]{->}(4.18,-0.6139148)(4.09,-0.5789148)(4.025,-0.5239148)(3.895,-0.378914794921875)
\rput[bl](4.235,-1.0389148){\normalsize{$r_1$}}
\rput[bl](4.29,-0.7139148){\normalsize{$r_\text{es}$}}
\psellipse[linecolor=black, linewidth=0.04, dimen=outer](7.6,0.041085206)(2.0,1.2)
\pscircle[linecolor=black, linewidth=0.04, dimen=outer](7.6,0.041085206){1.2}
\psline[linecolor=black, linewidth=0.04, arrowsize=0.05291667cm 2.0,arrowlength=1.4,arrowinset=0.0]{->}(10.735,-1.2239147)(9.16,-0.2839148)
\psline[linecolor=black, linewidth=0.04, arrowsize=0.05291667cm 2.0,arrowlength=1.4,arrowinset=0.0]{->}(9.18,0.2710852)(10.85,1.0960852)
\psline[linecolor=black, linewidth=0.04, arrowsize=0.05291667cm 2.0,arrowlength=1.4,arrowinset=0.0]{->}(9.005,0.021085205)(7.91,0.2810852)
\rput[bl](10.125,-0.7189148){\normalsize{$E_\text{in}$}}
\rput[bl](10.135,0.2910852){\normalsize{$E_\text{out}$}}
\rput[bl](7.74,0.44108522){\normalsize{$\Delta E$}}
\end{pspicture}
}
\caption{Schematic picture of the ergosphere (left) and the Penrose process (right).}
\label{figergo}
\end{figure}

In the Kerr geometry, the two vector fields~$\partial_t$ and~$\partial_\varphi$ are Killing fields.
The corresponding conserved quantities are
\begin{align}
E &:= \big\la \frac{\partial}{\partial t}, \dot{\gamma}(\tau) \big\ra_{\gamma(\tau)} &&
\hspace*{-2cm} \text{energy} \label{Edef} \\
A &:= \big\la \frac{\partial}{\partial \varphi}, \dot{\gamma}(\tau) \big\ra_{\gamma(\tau)} &&
\hspace*{-2cm} \text{angular momentum} \:. \label{Adef}
\end{align}
Let us consider the energy in more detail for a test particle moving along the geodesic~$\gamma$.
In this case, $\gamma(\tau)$ is a causal curve (i.e.\ $\dot{\gamma}(\tau)$ is timelike or
null everywhere), and we always choose the parametrization such that~$\gamma$ is future-directed
(i.e.\ the time coordinate~$\gamma^0(\tau)$ is monotone increasing in~$\tau$).
In the asymptotic end (i.e.\ for large~$r$), the Killing field~$\partial_t$ is timelike and
future-directed. As a consequence, the inner product in~\eqref{Edef} is strictly positive.
This corresponds to the usual concept of the energy being a non-negative quantity.
We point out that this result relies on the assumption that the Killing field~$\partial_t$ is timelike.
However, if this Killing field is spacelike, then the inner product in~\eqref{Edef} could very well
be negative. In order to verify if this case occurs, we compute
\begin{align*}
\big\la \frac{\partial}{\partial t},  \frac{\partial}{\partial t} \big\ra
&= g_{00} = \frac{\Delta}{U} - \frac{a^2 \sin^{2}\vartheta}{U} 
= \frac{1}{U} \Big( r^2 - 2 M r + a^2 \cos^2 \vartheta \Big) \:,
\end{align*}
where we read off the corresponding metric coefficient in~\eqref{BL}
and simplified it using~\eqref{UDdef}. Computing the roots, one sees that the
Killing field~$\partial_t$ indeed becomes null on the surface
\beq \label{ergosphere}
r=r_\text{es} := M +  \sqrt{M^2 - a^2\, \cos^2 \vartheta} \:,
\eeq
the so-called {\em{ergosphere}}. 
Comparing with the formula for the event horizon~\eqref{EH}, one sees that 
the ergosphere is outside the event horizon and intersects the event horizon at the
poles $\vartheta=0,\,\pi$ (see the left of Figure~\ref{figergo}).
The region~$r_1 < r < r_\text{es}$ is the so-called {\em{ergoregion}}.

The ergosphere causes major difficulties in the proof of linear stability of the Kerr geometry.
These difficulties are not merely technical, but they are related to physical phenomena,
as we now explain step by step.
The name ergosphere is motivated from the fact that it gives rise to a mechanism for extracting
energy from a rotating black hole. This effect was first observed by Roger Penrose~\cite{penrose-process}
and is therefore referred to as the {\em{Penrose process}}.
In order to explain this effect, we consider a spaceship of energy~$E_\text{in}$ which flies into
the ergoregion (see the right of Figure~\ref{figergo}), where it ejects a projectile of energy~$\Delta E$
which falls into the black hole. After that, the spaceship flies out of the ergoregion with
energy~$E_\text{out}$. Due to energy conservation, we know that~$E_\text{in} = E_\text{out} + \Delta E$.
By choosing the momentum of the projectile appropriately, one can arrange
that the energy~$\Delta E$ is negative. Then the final energy~$E_\text{out}$ is larger than the
initial energy~$E_\text{in}$, which means that we gained energy.
This energy gain does not contradict total energy conservation, because one should think of the
energy as being extracted from the black hole (this could indeed be made precise by
taking into account the back reaction of the space ship onto the black hole, but we do not have
time for entering such computations).
Therefore, the Penrose process is similar to the so-called ``swing-by'' or ``gravitational slingshot,''
where a satellite flies close to a planet of our solar system and uses the kinetic energy
of the planet for its own acceleration. The surprising effect is that in the Penrose process, one can extract energy
from the black hole, although the matter of the black hole is trapped behind the event horizon.

The wave analogue of the Penrose process is called {\em{superradiance}}.
Instead of the spaceship one considers a wave packet flying in the direction of the black hole.
The wave propagates as described by a corresponding wave equation (we will see such
wave equations in more detail later). As a consequence, part of the wave will ``fall into'' the black hole,
whereas the remainder will pass the black hole and will eventually leave the black hole region.
If the energy of the outgoing wave is larger than the energy of the oncoming wave, then
one speaks of superradiant scattering. This effect is quite similar to the Penrose process.
However, one major difference is that, in contrast to the Penrose process, 
there is no freedom in choosing the momentum of the projectile.
Instead, the dynamics is determined completely by the initial data, so that the only freedom is to
prepare the incoming wave packet. As we shall see later in this lecture, superradiance
indeed occurs for scalar waves in the Kerr geometry.

\section{The Scalar Wave Equation in the Kerr Geometry} \label{secscalar}
\begin{figure}
%
\psscalebox{1.0 1.0} 
{
\begin{pspicture}(-3.3,-1.306905)(11.618133,1.306905)
\definecolor{colour0}{rgb}{0.8,0.8,0.8}
\pspolygon[linecolor=colour0, linewidth=0.02, fillstyle=solid,fillcolor=colour0](0.013132858,0.94309497)(0.023132859,-0.726905)(0.40813285,-0.581905)(0.85813284,-0.451905)(1.2731328,-0.396905)(1.7281328,-0.386905)(2.2031329,-0.41690502)(2.7831328,-0.511905)(3.3581328,-0.691905)(4.013133,-0.891905)(4.588133,-1.006905)(5.323133,-1.011905)(5.843133,-0.941905)(6.438133,-0.816905)(7.188133,-0.586905)(7.178133,1.103095)(6.8331327,0.998095)(6.3331327,0.883095)(5.713133,0.828095)(5.138133,0.828095)(4.4781327,0.858095)(3.7931328,0.928095)(2.998133,1.043095)(2.258133,1.153095)(1.8931328,1.198095)(1.3531328,1.218095)(0.89313287,1.188095)(0.31813285,1.053095)
\rput[bl](3.8231328,-1.306905){\normalsize{$r_1$}}
\rput[bl](7.5181327,-0.661905){\normalsize{$\scrN_1$}}
\psline[linecolor=black, linewidth=0.04, arrowsize=0.05291667cm 2.0,arrowlength=1.4,arrowinset=0.0]{->}(3.7131329,-0.831905)(3.8681328,-0.29190502)
\rput[bl](1.7981329,0.293095){\normalsize{$\Omega$}}
\psbezier[linecolor=black, linewidth=0.04](0.008132858,-0.746905)(0.8055816,-0.39151636)(1.4789963,-0.34330606)(2.2655346,-0.42690500259399417)(3.052073,-0.51050395)(3.9470007,-0.99151635)(4.7360477,-1.016905)(5.5250945,-1.0422937)(6.2382083,-0.93050396)(7.188133,-0.586905)
\psbezier[linecolor=black, linewidth=0.04](0.008132858,0.953095)(0.7967578,1.2584836)(1.4939438,1.246694)(2.1000543,1.1830949974060059)(2.7061646,1.1194961)(3.911223,0.87848365)(4.691539,0.853095)(5.471855,0.82770634)(6.243719,0.78449607)(7.1831326,1.128095)
\psline[linecolor=black, linewidth=0.04, arrowsize=0.05291667cm 2.0,arrowlength=1.4,arrowinset=0.0]{->}(4.8331327,0.828095)(4.878133,1.398095)
\rput[bl](7.4981327,0.988095){\normalsize{$\scrN_2$}}
\rput[bl](3.468133,-0.506905){\normalsize{$\nu$}}
\rput[bl](4.468133,1.133095){\normalsize{$\nu$}}
\end{pspicture}
}
\caption{Conservation law corresponding to a Killing symmetry.}
\label{figconserve}
\end{figure}
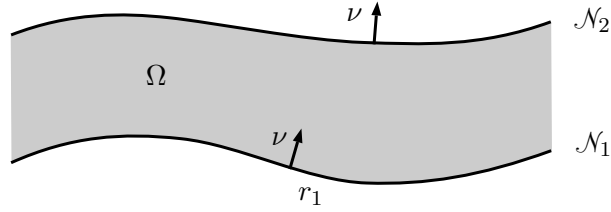

In preparation of the analysis of general linear wave equations, we begin with the simplest example:
the scalar wave equation. It has the useful property that it is of variational form, meaning that
it can be derived from an action principle. Indeed, choosing the Dirichlet action
\[ \Sact = \int_\scrM g^{ij}\: (\partial_i \phi)\, (\partial_j \phi)\: d\mu_\scrM \:, \]
(where~$d\mu_\scrM = \sqrt{|\det g|}\, d^4x$ is the volume measure induced by the Lorentzian metric),
demanding criticality for first variations gives the scalar wave equation
\[ 0 = \Box \phi := \nabla_i \nabla^i \phi \:. \]
The main advantage of the variational formulation is that Noether's theorem relates symmetries
to conservation laws. Another method for getting these conservation laws,
which is preferable to us because it is closely related to the notion of Killing fields, is to work directly with the
energy-momentum tensor of the field. Recall that in the Einstein equations~\eqref{einstein},
the Einstein tensor on the left is divergence-free as a consequence of the second Bianchi identities.
Therefore, the energy-momentum tensor is also divergence-free,
\beq \label{Tconserve}
\nabla^i T_{ij} = 0 \:.
\eeq
Now let~$K$ be a Killing field. Contracting the energy-momentum tensor with the Killing field
gives a vector field,
\[ u^i := T^{ij} \,K_j \:. \]
The calculation
\[ \nabla_i u^i = \big( \nabla_i T^{ij} \big)\: K_j + T^{ij} \, \nabla_i K_j = 0 \]
(where the first summand vanishes according to the conservation law~\eqref{Tconserve}, whereas the 
second summand is zero in view of the Killing equation~\eqref{killing} and the symmetry of the
energy-momentum tensor)
shows that this vector field is divergence-free.
Therefore, integrating the divergence of~$u$ over a space-time region~$\Omega$
and using the Gau{\ss} divergence theorem, we conclude that the flux integral of~$u$
through the surface~$\partial \Omega$ vanishes. The situation we have in mind is that the set~$\Omega$
is the region between two spacelike hypersurfaces~$\scrN_1$ and~$\scrN_2$
(see Figure~\ref{figconserve}).
Assuming that the vector field~$u$ has suitable decay properties at spatial infinity
(in the simplest case that it has spatially compact support), we obtain the conservation law
\beq \label{Tijconserve}
0 = \int_\Omega \nabla_i u^i \: d\mu_\scrM 
= \int_{\scrN_1} T_{ij}\, \nu^i \, K^j\: d\mu_{\scrN_1} - 
\int_{\scrN_1} T_{ij}\, \nu^i \, K^j\: d\mu_{\scrN_2} \:,
\eeq
where~$\nu$ is the future-directed normal on~$\scrN_{1\!/\!2}$ and~$d\mu_{\scrN_{1\!/\!2}}$
is the volume measure corresponding to the induced Riemannian metric.

In the Kerr geometry in Boyer-Lindquist coordinates, the Dirichlet action takes the explicit form
\[ \Sact = {\int_{-\infty}^{\infty}}\, dt
\int_{r_1}^{\infty}dr\int_{-1}^{1}d(\cos
\vartheta) \int_{0}^{\pi}d\varphi\,\L(\phi,\nabla \phi) \]
with
\begin{align*}
\L(\phi,\nabla \phi) &= -\Delta|\partial_{r}\phi|^{2}+\frac{1}{\Delta} \left|((r^{2}+a^{2})\partial_{t}
+a\partial_{\varphi})\phi \right|^{2} \nonumber \\
&\quad\; -\sin^{2}{\vartheta} \left|\partial_{\cos \vartheta}\phi \right|^{2} -\frac{1}{\sin^{2}\vartheta} \left| (a\sin^{2}\vartheta\partial_{t}+\partial_{\varphi})\phi \right|^{2} \:.
\end{align*}
Considering first variations, the scalar wave equation becomes
\beq \label{weq}
\begin{split}
& \left[  \frac{\partial}{\partial r} \Delta \frac{\partial}{\partial r} - \frac{1}{\Delta}
\left\{ (r^2+a^2)\: \frac{\partial}{\partial t} + a\: \frac{\partial}{\partial \varphi} \right\}^2 \right.
 \\
&\qquad \left. + \frac{\partial}{\partial \cos \vartheta} \:\sin^2 \vartheta\: \frac{\partial}{\partial \cos \vartheta} 
+ \frac{1}{\sin^2 \vartheta} \left\{ a \sin^2 \vartheta\: \frac{\partial}{\partial t} 
+ \frac{\partial}{\partial \varphi}  \right\}^2 \right] \phi = 0 \:.
\end{split}
\eeq
Using the formula for the energy-momentum tensor
\[ T_{ij} = (\partial_i \phi) (\partial_j \phi) - \frac{1}{2}\: (\partial_k \phi)\,(\partial^k \phi)\: g_{ij} \:, \]
the conserved energy becomes
\begin{align}
E &:= \int_{\scrN_t} T_{ij} \,\nu^j\, (\partial_t)^j \:d\mu_{\scrN_t}
= \int_{\scrN_t} T_{i0} \, (\partial_t)^j \:d\mu_{\scrN_t} \label{energygen} \\
&= \int_{r_1}^{\infty} dr \int_{-1}^{1} d(\cos\vartheta)
\int_{0}^{2\pi} d\varphi \:\mathcal{E} 
\end{align}
with the ``energy density''
\begin{align*}
{\mathcal{E}} &= 
\left({\frac{(r^{2}+a^{2})^{2}}{\Delta}} - a^{2}\,\sin^{2}\vartheta \right) \left|\partial_{t} \phi \right|^{2}+\Delta
\left|\partial_{r}\phi \right|^{2} \nonumber \\
&\quad\; +\sin^{2}\vartheta \left|\partial_{\cos
\vartheta}\phi \right|^{2}+\left(
{\frac{1}{\sin^{2}\vartheta}}-{\frac{a^{2}}{\Delta}}\right) \left| \partial_{\varphi}\phi \right|^{2} \:.
\end{align*}
Using~\eqref{UDdef}, one sees that the factor in front of the term~$|\partial_{\varphi}\phi|$ is everywhere
positive. However, the factor in front of the term~$|\partial_t \phi|^2$ is negative
precisely inside the ergosphere~\eqref{ergosphere}.
This consideration shows that, exactly as for point particles~\eqref{Edef}, the
energy of scalar waves may again be negative inside the ergosphere.

What does the indefiniteness of the energy tell us?
We first point out that it does {\em{not}} imply that superradiance really occurs, because in order to 
analyze superradiance, one must study the dynamics of waves.
Instead, it only means that there is a possibility for superradiance to occur.
In technical terms, the indefiniteness of the energy leads to the difficulty that energy conservation does not
give us control of the Sobolev norm of the wave. A possible scenario, which does not contradict
energy conservation, is that the amplitude of the wave
grows in time both inside and outside the the ergosphere.
It is a major task in proving linear stability to rule out this scenario.

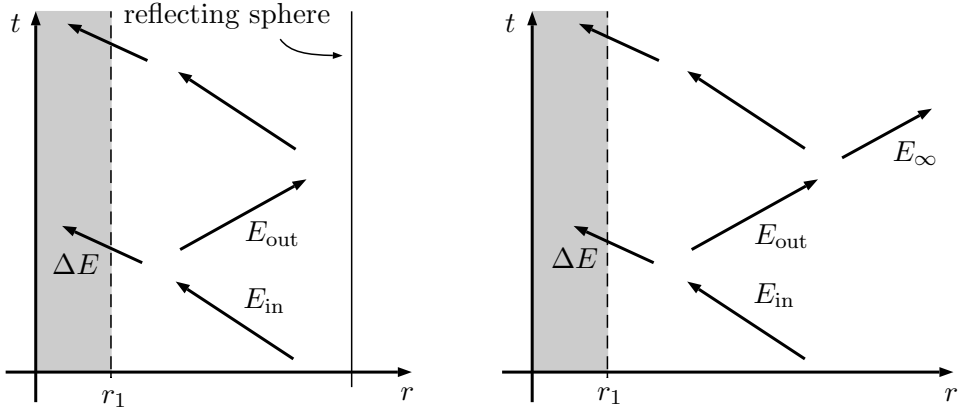
\begin{figure}
%
\psscalebox{1.0 1.0} 
{
\begin{pspicture}(-0.6,-2.675)(14.79,2.675)
\definecolor{colour0}{rgb}{0.8,0.8,0.8}
\psframe[linecolor=colour0, linewidth=0.02, fillstyle=solid,fillcolor=colour0, dimen=outer](8.01,2.6)(6.995,-2.195)
\psframe[linecolor=colour0, linewidth=0.02, fillstyle=solid,fillcolor=colour0, dimen=outer](1.41,2.6)(0.395,-2.195)
\rput[bl](5.25,-2.575){\normalsize{$r$}}
\rput[bl](1.275,-2.675){\normalsize{$r_1$}}
\psline[linecolor=black, linewidth=0.04, arrowsize=0.05291667cm 2.0,arrowlength=1.4,arrowinset=0.0]{->}(2.32,-0.575)(4.0,0.36)
\psline[linecolor=black, linewidth=0.04, arrowsize=0.05291667cm 2.0,arrowlength=1.4,arrowinset=0.0]{->}(0.4,-2.605)(0.4,2.595)
\psline[linecolor=black, linewidth=0.04, arrowsize=0.05291667cm 2.0,arrowlength=1.4,arrowinset=0.0]{->}(0.0,-2.205)(5.4,-2.205)
\psline[linecolor=black, linewidth=0.04, arrowsize=0.05291667cm 2.0,arrowlength=1.4,arrowinset=0.0]{->}(7.0,-2.605)(7.0,2.595)
\psline[linecolor=black, linewidth=0.04, arrowsize=0.05291667cm 2.0,arrowlength=1.4,arrowinset=0.0]{->}(6.6,-2.205)(12.6,-2.205)
\psbezier[linecolor=black, linewidth=0.02, linestyle=dashed, dash=0.17638889cm 0.10583334cm](1.4,2.595)(1.4,1.8565384)(1.4,-2.9434617)(1.4,-2.205)
\psline[linecolor=black, linewidth=0.02](4.6,2.595)(4.6,-2.405)
\psbezier[linecolor=black, linewidth=0.02, linestyle=dashed, dash=0.17638889cm 0.10583334cm](8.0,2.595)(8.0,1.8565384)(8.0,-2.9434617)(8.0,-2.205)
\rput[bl](7.865,-2.635){\normalsize{$r_1$}}
\rput[bl](12.48,-2.595){\normalsize{$r$}}
\rput[bl](0.065,2.325){\normalsize{$t$}}
\rput[bl](6.64,2.325){\normalsize{$t$}}
\psline[linecolor=black, linewidth=0.04, arrowsize=0.05291667cm 2.0,arrowlength=1.4,arrowinset=0.0]{->}(1.815,-0.75)(0.75,-0.26)
\psline[linecolor=black, linewidth=0.04, arrowsize=0.05291667cm 2.0,arrowlength=1.4,arrowinset=0.0]{->}(3.825,-2.03)(2.26,-1.0)
\psline[linecolor=black, linewidth=0.04, arrowsize=0.05291667cm 2.0,arrowlength=1.4,arrowinset=0.0]{->}(3.855,0.76)(2.29,1.79)
\psline[linecolor=black, linewidth=0.04, arrowsize=0.05291667cm 2.0,arrowlength=1.4,arrowinset=0.0]{->}(1.885,1.94)(0.82,2.43)
\psline[linecolor=black, linewidth=0.04, arrowsize=0.05291667cm 2.0,arrowlength=1.4,arrowinset=0.0]{->}(10.625,0.77)(9.06,1.8)
\psline[linecolor=black, linewidth=0.04, arrowsize=0.05291667cm 2.0,arrowlength=1.4,arrowinset=0.0]{->}(10.625,-2.03)(9.06,-1.0)
\psline[linecolor=black, linewidth=0.04, arrowsize=0.05291667cm 2.0,arrowlength=1.4,arrowinset=0.0]{->}(9.12,-0.575)(10.8,0.36)
\psline[linecolor=black, linewidth=0.04, arrowsize=0.05291667cm 2.0,arrowlength=1.4,arrowinset=0.0]{->}(8.615,-0.75)(7.55,-0.26)
\psline[linecolor=black, linewidth=0.04, arrowsize=0.05291667cm 2.0,arrowlength=1.4,arrowinset=0.0]{->}(8.685,1.94)(7.62,2.43)
\psline[linecolor=black, linewidth=0.04, arrowsize=0.05291667cm 2.0,arrowlength=1.4,arrowinset=0.0]{->}(11.12,0.645)(12.315,1.3)
\rput[bl](1.57,2.365){\normalsize{reflecting sphere}}
\psbezier[linecolor=black, linewidth=0.02, arrowsize=0.05291667cm 2.0,arrowlength=1.4,arrowinset=0.0]{->}(3.65,2.225)(3.78,2.045)(4.02,2.015)(4.48,2.005)
\rput[bl](0.62,-0.895){\normalsize{$\Delta E$}}
\rput[bl](7.25,-0.815){\normalsize{$\Delta E$}}
\rput[bl](3.16,-1.415){\normalsize{$E_\text{in}$}}
\rput[bl](3.18,-0.525){\normalsize{$E_\text{out}$}}
\rput[bl](9.94,-1.345){\normalsize{$E_\text{in}$}}
\rput[bl](9.95,-0.555){\normalsize{$E_\text{out}$}}
\rput[bl](11.795,0.545){\normalsize{$E_\infty$}}
\end{pspicture}
}
\caption{The black hole bomb (left) and wave propagation in the Kerr geometry (right).}
\label{figbomb}
\end{figure}

The basic difficulty can be understood qualitatively in more detail in the scenario of the
so-called {\em{black hole bomb}} as introduced by Press and Teukolsky~\cite{press-bomb}
and studied by Cardoso et al~\cite{cardoso}.
In this gedanken experiment, one puts a metal sphere around a Kerr black hole (as shown
schematically on the left of Figure~\ref{figbomb}.
We consider a wave packet of energy~$E_\text{in}$ inside the sphere flying towards the black hole.
Part of the wave will cross the event horizon, while the remainder will pass the black hole.
As in the above description of superradiance, we assume that the energy~$\Delta E$
of the wave crossing the event horizon is negative. Then the energy~$E_\text{out}$ of the
outgoing wave is larger than the energy~$E_\text{in}$ of the incoming wave.
The outgoing wave is reflected on the metal sphere, becoming a new wave which again
flies towards the black hole. If it can be arranged that the new incoming wave has the same
shape as the original wave, this process repeats itself, generating in each step
a certain positive energy. In this scenario, the energy density inside the metal sphere
would grow exponentially fast in time. When the energy density gets too large, the metal
sphere would explode, explaining the name ``black hole bomb.'' For clarity, we point out that
in this mechanism one always assumes that the total energy extracted from the black hole
is much smaller than the total rotational energy of the black hole, so that the back reaction
on the black hole need not be taken into account.

The black hole bomb suggests that, putting a metal sphere around the black hole
could lead to an instability, which would become manifest in an explosion of the metal sphere.
The point of interest in connection to the stability problem for rotating black holes is that
a very similar scenario might occur even without the metal sphere:
We again consider a wave packet flying towards the black hole. Again, part of the
wave with energy~$\Delta E$ crosses the event horizon, whereas the remainder of
energy~$E_\text{out}$ passes the black hole. The point is that
only part of this wave will reach null infinity. Another part will be backscattered by the
gravitational field and will again fly towards the black hole.
Therefore, except for the ``energy loss'' $E_\infty$ by the part of the wave propagating
to null infinity, we are again in the scenario of the black hole bomb where the process
repeats itself, potentially leading to an exponential increase in time of the amplitude of the wave.

Clearly, this picture is oversimplified because, instead of wave packets, one must consider
waves which are spread out in space, leading to a nonlocal problem.
Nevertheless, in this picture it becomes clear why the problem of linear stability of rotating
black holes amounts to a quantitative question:
Can the initial wave packet be arranged such that
the ``energy gain''~$-\Delta E$ is larger than the ``energy loss'' $E_\infty$?
If the answer is yes, a rotating black hole should be unstable, and it should decay
by radiation of gravitational waves to a Schwarzschild black hole.
It is the main goal of these lectures to explain why this does {\em{not}} happen, i.e.\
why rotating black holes are linearly stable.
Before we can enter this problem in mathematical detail, we need to introduce
linear wave equations and review a few structural results.

\section{An Overview of Linear Wave Equations in the Kerr Geometry} \label{secoverview}
\subsection{The Dirac Equation} \label{secdirac}
In these lectures I shall not enter the details of the Dirac equation, although
most of my work has been concerned with or related to the Dirac equation.
I only want to explain why the analysis for the Dirac equation
is much {\em{easier}} than for other wave equations.

The Dirac equation describes a relativistic quantum mechanical particle with spin.
In order to keep the setting as simple as possible, we work in coordinates and
local trivializations of the spinor bundle (which has the advantage that we do not need to
even define what the spinor bundle is). Then the Dirac wave function~$\psi(x) \in \C^4$
has four complex components, which describe the spinorial degrees of freedom of the wave function.
The Dirac equation reads
\[ \big(i \gamma^j(x)\, \partial_j + \B - m\big) \psi = 0 \:. \]
Here~$m$ is the rest mass of the Dirac particles, and the four matrices~$\gamma^j$ encode the
Lorentzian metric via the anti-commutation relations
\[ \big\{ \gamma^j(x), \gamma^k(x) \big\} = 2 g^{jk}(x) \,\1_{\C^4} \:, \]
where the anti-commutator is defined by
\[ \big\{ \gamma^j, \gamma^k \big\} := \gamma^j\, \gamma^k + \gamma^k\, \gamma^j\: \:. \]
The multiplication operator~$\B$ involves the so-called spin coefficients,
which, in analogy to the Christoffel symbols of the Levi-Civita connection,
are formed of first partial derivatives of the Dirac matrices.
We do not need the details here and refer instead to the explicit formulas
in~\cite{u22} or~\cite[Chapter~3]{intro}.

Coming from quantum mechanics, the Dirac equation has additional structures
which allow for the probabilistic interpretation of the wave function.
In particular, there is a quantity which can be interpreted as the
probability density as seen by an observer, and the spatial integral of this probability
density is equal to one, for any fixed time of the observer.
This probability integral is described mathematically as follows.
The spinors at a space-time point~$x \in \scrM$ are endowed with
an indefinite inner product of signature~$(2,2)$, which we
denote by~$\Sl \psi | \psi \Sr_x$. For any solution~$\psi$ of the Dirac equation,
the pointwise expectation value of the Dirac matrices with respect to this inner product
defines a vector field
\[ J^k(x) := \Sl \psi(x) \,|\, \gamma^k \: \psi(x) \Sr_x \:. \]
This vector field is the so-called {\em{Dirac current}}.
The structure of the Dirac equation ensures that this vector field is always
non-spacelike and future-directed. Moreover, as a consequence of the Dirac equation, this
vector field is divergence-free, i.e.\
\[ \nabla_k J^k(x) = 0 \]
(where~$\nabla$ is again the Levi-Civita connection); this is referred to as {\em{current conservation}}.
Integrating this equation over a region~$\Omega$ between two spacelike hypersurfaces
(as shown in Figure~\ref{figconserve}), one obtains the conservation law
\beq \label{curcons}
\int_{\scrN_2} J_i\, \nu^i \, d\mu_{\scrN_2} = 
\int_{\scrN_1} J_i\, \nu^i \, d\mu_{\scrN_1}
\eeq
(here we again assume that the Dirac wave function has suitable decay properties at
spatial infinity). In view of this conservation law and the linearity of the Dirac equation,
one can normalize the Dirac solutions such that the integral in~\eqref{curcons} equals one.
Then the integrand in~\eqref{curcons} has the interpretation as the probability density
for an observer for whom the spacelike hypersurface~$\scrN_1$ (or~$\scrN_2$)
describes space. We point out that the probability density is non-negative
simply as a consequence of the fact that the current vector is non-spacelike
and future-directed, and that the normal is timelike and future-directed.
In particular, the probability density is non-negative even inside the ergosphere;
note also that, in contrast to~\eqref{Tijconserve}, the integrand in~\eqref{curcons} does not
involve a Killing field.

These structures coming from the probabilistic interpretation of the Dirac equation
are a major simplification for analyzing the long-time dynamics of Dirac waves
in the Kerr geometry. Namely, the current integral~\eqref{curcons} can be used to
define a scalar product on the solutions of the Dirac equation by
\[ (\psi | \phi)_t := \int_{N_t} \Sl \psi | \gamma^j \phi \Sr_x\: \nu_j\: d\mu_{\scrN_t} \:, \]
where~$\scrN_t$ is the surface of constant time in Boyer-Lindquist coordinates
outside the event horizon, and where we restrict attention to Dirac solutions with suitable
decay properties on the event horizon and near spatial infinity (for example
wave functions with spatially compact support outside the event horizon).
Taking the completion, we obtain the Hilbert space~$(\H_t, (.|.)_t)$ of Dirac solutions.
The conservation law~\eqref{curcons} means that the time evolution
operator~$U_{t,t_0} : \H_{t_0} \rightarrow \H_t$
from time~$t_0$ to time~$t$ is a unitary operator. Since the Kerr geometry is stationary,
we can canonically identify the Hilbert space~$\H_t$ with~$\H_{t_0}$ by 
time translation of the wave functions.
Moreover, the unitary time evolution can be written as
\beq \label{spectralunit}
U_{t,t_0} = e^{-i (t-t_0) H} \:,
\eeq
where the so-called Dirac Hamiltonian~$H$ is a self-adjoint operator on the Hilbert space
(the self-adjointness extension can be constructed in general using Stone's theorem
or Chernoff's method~\cite{chernoff73}).
In this way, the long-time dynamics can be related to spectral properties of a
{\em{self-adjoint operator on a Hilbert space}}. No superradiance phenomena occur.

More details on the Dirac equation and the above method
can be found in my joint papers with Niky Kamran, Joel Smoller and Shing-Tung
Yau~\cite{kerr, tkerr, decay}. For the general method of constructing self-adjoint extensions,
the more recent paper~\cite{chernoff} may be useful.

\subsection{Massless Equations of General Spin, the Teukolsky Equation}
After the short excursion to quantum mechanics, we now return to {\em{classical waves}}.
The waves of interest are
\[ \left\{ \begin{array}{c} \text{scalar waves} \\ \text{electromagnetic waves} \\ \text{gravitational waves} \:.
\end{array} \right. \]
Scalar waves were already considered in Section~\ref{secscalar};
they are studied mainly because of their mathematical simplicity.
The waves of physical interest are electromagnetic and gravitational waves.
Note that all these waves are {\em{massless}}.

All the above wave equations can be described in a unified framework
due to Teukolsky~\cite{teukolsky}. We now explain schematically how the
Teukolsky formulation works. Since the involved computations are quite lengthy,
we cannot enter the details but refer instead to the textbook~\cite{chandra}.
The Teukolsky equation is derived in the {\em{Newman-Penrose formalism}}, which we
now briefly introduce. In this formalism, one works with a
{\em{double null frame}}, i.e.\ with a set of vectors~$(l,n,m,\overline{m})$
of the complexified tangent space with inner products
\[ \la l,n \ra = 1 \:,\qquad \la m, \overline{m} \ra = -1 \:, \]
whereas all other inner products vanish,
\[ \la l,l \ra = \la n,n \ra =  \la m,m \ra = \la \overline{m}, \overline{m} \ra = 0 \:. \]
In the example of Minkowski space in Cartesian coordinates~$(t,x,y,z)$, one can choose~$l$ and~$n$
as two null vectors, for example
\[ l = \frac{1}{\sqrt{2}}\: \Big( \frac{\partial}{\partial t} + \frac{\partial}{\partial x} \Big) \qquad \text{and} \qquad
n = \frac{1}{\sqrt{2}}\: \Big( \frac{\partial}{\partial t} - \frac{\partial}{\partial x} \Big) \:. \]
The orthogonal complement of these two vectors is the two-dimensional spacelike
plane spanned by the vectors~$\partial_y$ and~$\partial_z$. Therefore, the only way
to obtain additional null vectors is to {\em{complexify}} by choosing for example
\[ m = \frac{1}{\sqrt{2}}\: \Big( \frac{\partial}{\partial y} + i \frac{\partial}{\partial z} \Big) \qquad \text{and} \qquad
\overline{m} = \frac{1}{\sqrt{2}}\: \Big( \frac{\partial}{\partial y} - i \frac{\partial}{\partial z} \Big) \:. \]
Likewise, on a Lorentzian manifold, the vectors~$(l,n,m,\overline{m})$ form a basis
of the complexified tangent space.
The Lorentzian inner product~$\la .,. \ra$ is extended
to the complexified tangent space as a bilinear form (not sesquilinear; thus
no complex conjugation is involved).
The double null frame is well-suited for the analysis of the vacuum Einstein equations
(indeed, the Kerr solution was discovered in the Newman-Penrose formalism).

Next, one combines the tensor components in the double null frame in
complex-valued functions. In the example of the {\em{Maxwell field}}, this works as follows.
The electromagnetic field tensor~$F_{ij}$ has six real components (three electric and three
magnetic field components). One combines these six real components to
the three complex functions
\beq \label{Psimax}
\Psi_0 = F_{lm}\:,\qquad \Psi_1 = \frac{1}{2} \,\big( F_{ln} + F_{m \overline{m}} \big) \:,\qquad
\Psi_2 = F_{n \overline{m}} \:.
\eeq
Then the homogeneous Maxwell equations
\[ dF = 0 \qquad \text{and} \qquad \nabla^k F_{jk} = 0 \]
give rise to a first-order system of partial differential equations for~$\Psi=(\Psi_0, \Psi_1, \Psi_2)$.
For the {\em{gravitational field}}, one considers similarly the Weyl tensor~$C_{ijkl}$.
Linearized gravitational waves are described by linear perturbations of
the Weyl tensor in the Newman-Penrose frame. Denoting the linear perturbation
of the Weyl tensor by~$W$, its ten real components are combined to the five complex functions
\begin{gather*}
\Psi_0 = - W_{lmlm}\:,\qquad \Psi_1 = -W_{lnlm}\:,\qquad \Psi_2 = -W_{lm\overline{m}n} \\
\Psi_3 = -W_{ln\overline{m}n}\:,\quad \Psi_4 = -W_{l\overline{m}n\overline{m}}\:.
\end{gather*}
Linearizing the second Bianchi identities~$R_{ij (kl; m)}=0$ gives a first-order system
of partial differential equations for~$\Psi=(\Psi_0, \ldots, \Psi_4)$.
In this formulation, the connection to the {\em{spin}} can be obtained simply by counting
the number of degrees of freedom. In quantum mechanics, a wave function of spin~$s$ has~$2s+1$
complex components. Therefore, we obtain the correct number of degrees of freedom if we set
\[ \begin{array}{rl} \text{electromagnetic waves:} 
&\quad \text{spin~$s=1$} \\ \text{gravitational waves:} &\quad \text{spin~$s=2$} \:.
\end{array}  \]
We remark that the connection to the spin is more profound than merely counting the number of degrees
of freedom, but we have no time to explain how this works.
For our purposes, it suffices to take the spin~$s$ as a parameter which characterizes the massless wave equation
by counting the number of components of the Newman-Penrose wave function~$\Psi
= (\Psi_0, \ldots, \Psi_{2s})$.

We write the first-order system of partial differential equations for electromagnetic waves
or linearized gravitational waves  symbolically as
\beq \label{system}
\D \begin{pmatrix} \Psi_0 \\ \vdots \\ \Psi_{2s} \end{pmatrix} = 0 \:.
\eeq
Working with this first-order system is not convenient for larger spin because the number
of equations gets large, and the equations are coupled in a complicated way.
But, as discovered by Teukolsky, the system of equations can be decoupled such as
to obtain a second-order partial differential equation for one complex-valued function.
This {\em{decoupling}} works schematically as follows:
One chooses a Newman-Penrose null frame
where~$l$ and~$n$ are aligned with the repeated principal null directions of the Weyl tensor
(in this frame, the Newman-Penrose components of the Weyl tensor
satisfy the equations~$\psi_0=\psi_1=\psi_3=\psi_4=0$, with~$\psi_2$ being the only non-zero component). 
Multiplying the linear first-order system~\eqref{system} in this frame by
a suitable first-order differential operator~$D$, we obtain the equation
\[ 0 = D {\mathcal{D}} \begin{pmatrix} \Psi_0 \\ \vdots \\ \Psi_{2s} \end{pmatrix}
= \left(\!\! \begin{array}{ccccc} T_0 & {0} & \cdots & {0} & {0} \\
* & * & \cdots & * & * \\
\vdots & \vdots & \ddots & \vdots & \vdots \\
* & * & \cdots & * & * \\
{0} & {0} & \cdots &{0} & {T_{2s}}
\end{array} \!\!\right) \begin{pmatrix} \Psi_0 \\ \Psi_1 \\ \vdots \\ \Psi_{2s-1} \\ \Psi_{2s} \end{pmatrix} \:, \]
where the stars stand for differential operators which we do not need to specify here.
The point is that this procedure generates zeros in the first and last row of the matrix,
giving rise to decoupled equations for the first and and last components of the Newman-Penrose
wave function~$\Psi$,
\beq \label{firstlast}
T_0\, \Psi_0 = 0 = T_{2s}\, \Psi_{2s} \:.
\eeq
Once the solution~$\Psi_0$ or~$\Psi_{2s}$ is known, all the other components of~$\Psi$
can be obtained by employing the so-called {\em{Teukolsky-Starobinsky identities}}, which have
similarities to the ``ladder operator'' for the harmonic oscillator used for obtaining
the excited states from the ground state. With this in mind, in what follows we restrict attention
to the equations for~$\Psi_0$ or~$\Psi_{2s}$ in~\eqref{firstlast}.
After detailed computations for the electromagnetic field
and for linearized gravitational fields, in both cases one ends up with the
same equation, except for a parameter~$s$ describing the spin.
We thus obtain the {\em{Teukolsky equation}} (sometimes called Teukolsky master equation;
we use the form of the equation as given in~\cite{whiting})
\beq \label{teukolsky}
\begin{split}
&\bigg( \frac{\partial}{\partial r} \Delta \frac{\partial}{\partial r} - \frac{1}{\Delta}
\left\{ (r^2+a^2)\: \frac{\partial}{\partial t} + a\: \frac{\partial}{\partial \varphi} - (r-M) \,s \right\}^2 
- 4 s \: (r+i a \cos \vartheta)\: \frac{\partial}{\partial t} \\
&\qquad + \frac{\partial}{\partial \cos \vartheta} \:\sin^2 \vartheta\: \frac{\partial}{\partial \cos \vartheta} 
+ \frac{1}{\sin^2 \vartheta} \left\{ a \sin^2 \vartheta\: \frac{\partial}{\partial t} 
+ \frac{\partial}{\partial \varphi} + i s \cos \vartheta \right\}^2 \bigg) \phi = 0 \:.
\end{split}
\eeq
For~$s=1$, this equation describes the first component~$\Psi_0$ of the Newman-Penrose
wave function~$(\Psi_0, \Psi_1, \Psi_2)$ for electromagnetic waves,
whereas the parameter value~$s=-1$ gives the equation for~$\Psi_2$.
Likewise, setting~$s=2$ gives the first component~$\Psi_0$ of the
Newman-Penrose wave function~$(\Psi_0, \ldots, \Psi_4)$ for gravitational waves,
whereas~$s=-2$ gives the equation for~$\Psi_4$.
By direct inspection, one sees that setting~$s=0$ gives back the scalar wave equation~\eqref{weq}.
We remark that setting~$s=\frac{1}{2}$ gives the massless Dirac equation~\cite{teukolsky}, and~$s=\frac{3}{2}$
gives the massless Rarita-Schwinger equation~\cite{guven}.

We close with a remark on gravitational perturbations. As outlined above, our method is to
consider perturbations of the Weyl tensor.
Alternatively, one could consider perturbations of the metric (indeed, this
was historically the first approach, going back to the stability analysis by Regge and Wheeler~\cite{reggestab}).
Working with {\em{metric perturbations}} has the disadvantage that infinitesimal coordinate transformations
also lead to perturbations of the metric, which however have no geometric significance.
In other words, when working with metric perturbations, the diffeomorphism invariance leads
to a gauge freedom which is not easy to handle. This was the original motivation for
Teukolsky and Press to consider instead perturbations of geometric quantities like the
Newman-Penrose components of the Weyl tensor, leading to the Teukolsky framework.
However, for some applications (for example in order to include matter models or
to describe nonlinear waves) it is necessary to work with metric perturbations.
Therefore, working in the Teukolsky formulation, the following question remains:
Given a linear perturbation of the Weyl tensor, how can it be realized by a metric perturbation?
This is an interesting and in general difficult question which we cannot analyze here
(see however~\cite{whiting+price} and the references therein).

\section{Separation of the Teukolsky Equation}
The Teukolsky equation~\eqref{teukolsky} has the remarkable property that it
can be completely separated into a system of ordinary differential equations (ODEs):
Due to the stationarity and axisymmetry, we can separate the
$t$- and $\varphi$-dependence with the usual plane-wave ansatz
\begin{equation}\label{separansatz}
\phi(t,r,\vartheta,\varphi) \;=\; e^{-i\omega
t-ik\varphi}\:\phi(r,\vartheta)\:,
\end{equation}
where $\omega$ is a quantum number which could be real or complex
and which corresponds to the ``energy'', and $k \in {\mathbb{Z}}/2$ is
a quantum number corresponding to the projection of angular momentum
onto the axis of symmetry of the black hole
(if~$s$ is half an odd integer, then so is~$k$).
Substituting~\eqref{separansatz} into~\eqref{teukolsky}, we see that the
Teukolsky operator splits into the sum of radial and angular parts,
giving rise to the equation
\[ 
({\mathcal{R}}_{\omega , k}+{\mathcal{A}}_{\omega, k}) \,\phi = 0 \:,
\] 
where ${\mathcal{R}}_{\omega , k}$ and ${\mathcal{A}}_{\omega , k}$
are given by (for details see~\cite[Section~6]{stable})
\begin{align}
{\mathcal{R}}_\omega &= -\frac{\partial}{\partial r} \Delta \frac{\partial}{\partial r} 
-\frac{1}{\Delta} \Big( \omega \,(r^2+a^2) + ak - i (r-M) \,s \Big)^2 - 4 i s r \omega + 4 k\, a \omega \label{Rop} \\
\A_\omega &= - \frac{\partial}{\partial \cos \vartheta} \:\sin^2 \vartheta\: \frac{\partial}{\partial \cos \vartheta} 
+\frac{1}{\sin^2 \vartheta} \Big( -a \omega \sin^2 \vartheta +k - s \cos \vartheta \Big)^2 \:. \label{Aop}
\end{align}

We can therefore separate the variables $r$ and $\vartheta$
with the multiplicative ansatz
\beq \label{lastsep}
\phi(r, \vartheta) \;=\; R(r)\: \Theta(\vartheta)\:,
\eeq
to obtain for given $\omega$ and $k$ the system of ODEs
\begin{equation} \label{coupled}
{\mathcal{R}}_{\omega ,k}\,R_\lambda \;=\; -\lambda \,
R_{\lambda},\qquad  {\mathcal{A}}_{\omega ,k}\,\Theta_\lambda \;=\;
\lambda \, \Theta_\lambda\:.
\end{equation}
Solutions of the coupled system~\eqref{coupled} are referred to as {\em{mode solutions}}.

We point out that the last separation~\eqref{lastsep} is not obvious
because it does not correspond to an underlying space-time symmetry.
Instead, as discovered by Carter for the scalar wave equation~\cite{carterrepub},
it corresponds to the fact that in the Kerr geometry there exists an irreducible quadratic Killing tensor
(i.e.\ a Killing tensor which is not a symmetrized tensor product of Killing vectors).
The separation constant $\lambda$ is an eigenvalue of the angular
operator ${\mathcal{A}}_{\omega, k}$ and can thus be thought of as
an angular quantum number. In the spherically symmetric
case~$a=0$, this separation constant goes over to the usual eigenvalues
$\lambda=l(l+1)$ of the total angular momentum operator.

\section{Results on Linear Stability and Superradiance}
Being familiar with the structure of the different linear wave equations,
we can now state our results on stability and superradiance.
The problem of {\em{linear stability}} of black holes amounts to the question
whether solutions of the corresponding linear wave equations decay for large times.
In order to put our results into context, we point out that the problem of
linear stability of black holes has a long history.
It goes back to the study of the Schwarzschild black hole by
Regge and Wheeler~\cite{reggestab} who showed
that an integral norm of the perturbation of each angular mode is bounded uniformly in time.
Decay of these perturbations was first proved in~\cite{friedmanstab}.
More detailed estimates of metric perturbations in Schwarzschild were
obtained in~\cite{dafermos-holzegel-rodnianski, hung+keller+wang}.
For the Kerr black hole, linear stability under perturbations of
general spin has been an open problem for many years,
which was solved in the dynamical setting in~\cite{stable}
(for related results obtained with different methods
see~\cite{metcalfe+tataru+tohaneanu2, andersson-blue1, andersson-baeckdahl-blue, dafermos-holzegel-rodnianski2}
and the references in these papers).
A key ingredient to our proof is the so-called {\em{mode stability}} result obtained by
Whiting~\cite{whiting}, who proved that the Teukolsky equation does not admit solutions which
decay both at spatial infinity and at the event horizon and increase exponentially in time.

We consider the Cauchy problem for the Teukolsky equation. Thus we seek a
solution~$\phi$ of the Teukolsky equation~\eqref{teukolsky} for given initial data
\[ \phi|_{t=0} = \phi_0 \qquad \text{and} \qquad \partial_t \phi|_{t=0} = \phi_1 \:. \]
Being a linear hyperbolic PDE, the Cauchy problem for the Teukolsky
equation has unique global solutions.
Also, taking smooth initial data, the solution is smooth for all times.
Our task is to show that solutions decay for large times.
In order to avoid specifying decay assumptions at the event horizon and at spatial infinity,
we restrict attention to compactly supported initial data outside the event horizon,
\beq \label{initsmooth}
\phi_0, \phi_1 \in C^\infty_0\big((r_1, \infty) \times S^2 \big) \:.
\eeq

Since the Kerr geometry is axisymmetric, the Teukolsky equation decouples into separate
equations for each azimuthal mode. Therefore, the solution of the Cauchy problem
is obtained by solving the Cauchy problem for each azimuthal mode and taking the sum of the resulting solutions.
With this in mind, we restrict attention to the Cauchy problem for a single azimuthal mode, i.e.
\beq \label{fouriermode}
\phi_0(r, \vartheta, \varphi) = e^{-i k \varphi}\: \phi_0^{(k)}(r, \vartheta)\:,\qquad
\phi_1(r, \vartheta, \varphi) = e^{-i k \varphi}\: \phi_1^{(k)}(r, \vartheta)
\eeq
for given~$k \in \Z/2$.
The main result of~\cite{stable} is stated as follows:
\begin{Thm} \label{thmdecay}
Consider a non-extreme Kerr black hole of mass~$M$ and angular momentum~$aM$
with~$M^2>a^2>0$. Then for any~$s \geq \frac{1}{2}$ and any~$k \in \Z/2$, the solution of the
Teukolsky equation with initial data of the form~\eqref{initsmooth} and~\eqref{fouriermode}
decays to zero in~$L^\infty_\text{\rm{loc}}((r_1, \infty) \times S^2)$.
\end{Thm} \noindent
This theorem establishes in the dynamical setting that the non-extreme Kerr black hole
is linearly stable.

Our method of proof uses an integral representation of the time evolution operator
involving the radial and angular solutions of the separated system of ODEs~\eqref{coupled}. Such an integral representation was derived earlier for
the {\em{scalar wave equation}} in~\cite{weq}, and it was used for proving decay in time~\cite{wdecay}.
Moreover, in~\cite{penrose} it was proven in the dynamical setting
that {\em{superradiance}} occurs for scalar waves. 
We now explain this result.
Superradiance for scalar waves in the Kerr geometry was first studied
by Zel'dovich and Starobinsky~\cite{zeldovich, starobinsky} on the level of modes.
More precisely, they computed the transmission and reflection coefficients
for the radial ODE in~\eqref{coupled}.
The absolute value squared of these coefficients can be interpreted as the
energy flux of the incoming and outgoing waves, respectively.
Comparing these fluxes, one obtains the relative energy gain.
Starobinsky computed the relative gain of energy to
about~$5\%$ for $k=1$ and less than~$1\%$ for $k \geq2$.

Unfortunately, this mode analysis does not give information on the dynamics.
Thus for a rigorous treatment of energy extraction one needs to consider the
time-dependent situation. In~\cite{penrose}, this is accomplished
by constructing initial data of the form of wave packets, in such a way that
the energy gain agrees with the results of the mode analysis up to an arbitrarily
small error. The crucial analytical ingredient to the proof
is the time-independent energy estimate for the outgoing wave as derived in~\cite{sobolev}. 

The remainder of these lectures is devoted to giving an outline of the proof of Theorem~\ref{thmdecay}.
Before entering the constructions, we point out the main difficulties:
\bitem
\itemD The Teukolsky equation for~$s \neq 0$ is {\em{not of variational form}}, i.e.\ it cannot
be obtained as the Euler-Lagrange equation of an action.
\itemD As a consequence, we cannot apply Noether's theorem to obtain
conserved quantities. In particular, there is {\em{no conserved energy}}, being an integral
of an energy density. This means that, in contrast to the situation described
for the Dirac equation in Section~\ref{secdirac}, the time evolution cannot be
described by a unitary operator on a Hilbert space.
As a consequence, we cannot use the spectral theorem for selfadjoint or unitary operators
on Hilbert spaces.
\itemD A related difficulty is that the {\em{coefficients}} of the first derivative terms
in the Teukolsky equation for~$s \neq 0$
are {\em{complex}}. Such complex potentials in a wave equation usually describe dissipation,
implying that (depending of the sign of the dissipation terms)
the solutions typically decay or increase exponentially in time.
This means that, in order to show that the solution of the Teukolsky equation
decays for large times, one must carefully control the signs and the size of the
complex coefficients by quantitative estimates.
\itemD In the separation of variables~\eqref{coupled},
both the radial and angular differential operators~${\mathcal{R}}_{\omega ,k}$
and~${\mathcal{A}}_{\omega ,k}$ depend on the separation constants~$k$ and~$\omega$.
As a consequence, it is not at all obvious if and how for given initial data
one can decompose the corresponding solution
of the Cauchy problem into a superposition of mode solutions.
An obvious difficulty is that, for such a {\em{mode decomposition}}, one would have to know the separation
constant~$\omega$, which in turn can be specified only if we already
know the full dynamics of the wave.
\eitem

\section{Hamiltonian Formulation and Integral Representations}
In order to analyze the dynamics of the Teukolsky wave, it is useful to work with contour integrals
of the resolvent of the Hamiltonian, as we now outline. In preparation, we must rewrite the Teukolsky equation in
Hamiltonian form. To this end, we introduce the two-component wave function
\[ \Psi = \sqrt{r^2+a^2} \begin{pmatrix} \phi \\ i \partial_t \phi \end{pmatrix} \]
and write the Teukolsky equation as
\beq \label{HPsi}
i \partial_t \Psi = H \Psi \:,
\eeq
where~$H$ is a second-order spatial differential operator.
We consider~$H$ as an operator on a Hilbert space~$\H$
with the domain
\[ \D(H) = C^\infty_0 \big( (r_1, \infty) \times S^2, \C^4 \big)\:. \]

It would be desirable to represent~$H$ as a self-adjoint operator on a Hilbert space~$\H$,
because it would then be possible to apply the spectral calculus and write the time
evolution operator similar as for the Dirac equation in the form~\eqref{spectralunit}.
Unfortunately, this procedure does {\em{not}} work here, as can be understood as follows.
As already mentioned at the end of the previous section, the Teukolsky equation is not
of variational form, implying that there is no conserved energy. If there were
a conserved bilinear form~$\la \Psi | \Phi \ra$ on the solutions, then the calculation
\[ 0 = \partial_t \la \Psi | \Phi \ra = \la \dot{\Psi} | \Phi \ra + \la \Psi | \dot{\Phi} \ra
= i \big( \la H \Psi | \Phi \ra - \la \Psi | H \Phi \ra \big) \]
would imply that the Hamiltonian were symmetric with respect to this bilinear form.
But, having no conserved energy, there is also no bilinear form with respect to which
the Hamiltonian is symmetric.
In order to avoid confusion, we remark that there is a conserved physical energy,
which in the example of a Maxwell field could be written in the form~\eqref{energygen}
with~$T_{ij}$ the energy-momentum tensor of the Maxwell field.
However, this energy involves all the components of the field tensor or, in other words,
all the components of the Newman-Penrose wave function in~\eqref{Psimax}.
Since the Teukolsky equation only gives~$\Psi_0$ or~$\Psi_2$, we would have to
compute the other components using the Teukolsky-Starobinsky identities.
As a consequence, the resulting formula for the Maxwell energy would involve
higher derivatives of the Teukolsky wave function, making the situation very
complicated. This is why we decided not to use the physical energy in our construction.

We conclude that we shall treat the operator~$H$ as a non-symmetric operator on
a Hilbert space. In order to get an idea for how to work with non-symmetric operators, it is helpful
get a motivation from the finite-dimensional setting. Thus let~$A$ be a linear operator
on a finite-dimensional Hilbert space~$\H$. Clearly, this operator need not be
diagonalizable, because Jordan chains may form. Nevertheless, one can get a spectral
calculus by working with contour integrals:
\begin{Lemma} Let~$A$ be a linear operator~$A$
on a Hilbert space~$\H$ of dimension~$n < \infty$. Then
\beq \label{CauchyA}
e^{-i t A} = -\frac{1}{2 \pi i} \ointctrclockwise_\Gamma e^{-i \omega t}\: \big(A-\omega)^{-1}\: d\omega\:,
\eeq
where~$\Gamma$ is a contour which encloses the whole spectrum of~$A$ with winding number one.
\end{Lemma}
\Proof If~$A$ is diagonalizable, we can choose a basis where~$A$ is diagonal,
\[ A = \text{diag} ( \lambda_1, \ldots, \lambda_n ) \:. \]
In this case, \eqref{CauchyA} is obtained immediately by carrying out the contour integral
for each matrix entry with the help of the Cauchy integral formula.

The case that~$A$ is not diagonalizable can be obtained by approximation, noting
that the diagonalizable matrices are dense and that both sides of~\eqref{CauchyA}
are continuous on the space of matrices (endowed with the topology of~$\C^{n \cdot n}$).
\QED \noindent
Motivated by this formula for matrices, we can hope that the Cauchy problem for the
equation~\eqref{HPsi} with initial data~$\Psi_0$ could be solved with the Cauchy integral formula by
\beq \label{Cauchy}
\Psi(t) = -\frac{1}{2 \pi i} \ointctrclockwise_\Gamma e^{-i \omega t}\: \big(H-\omega)^{-1}\: \Psi_0\: d\omega\:,
\eeq
where~$\Gamma$ is a contour which encloses all eigenvalues of~$H$
(note that this formula holds for any matrix~$H$, even if it is not diagonalizable).
It turns out that in our infinite-dimensional setting, this formula indeed holds.
The first step in making sense of this formula is to localize the spectrum of~$H$
and to make sure that the resolvent exists along the integration contour.
To this end, we choose the scalar product on~$\H$
as a suitable weighted Sobolev scalar product in such a way that
that the operator~$H-H^*$ is bounded, i.e.
\[ \| H - H^* \| \leq \frac{c}{2} \]
with a suitable constant~$c>0$.
Then we prove that the resolvent~$R_\omega:= (H - \omega)^{-1}$
exists if~$\omega$ lies outside a strip enclosing the real axis (see~\cite[Lemma~4.1]{stable}):
\begin{Lemma} \label{lemmaresex}
For every~$\omega$ with
\[ |\im \omega| > c \:, \]
the resolvent~$R_\omega = (H-\omega)^{-1}$ exists and is bounded by
\[ \| R_\omega \| \leq \frac{1}{|\im \omega|-c}\:. \]
\end{Lemma} \noindent
When forming contour integrals, one must always make sure to stay outside
the strip~$|\im \omega| \leq c$, making it impossible to work with closed contours enclosing the spectrum.
But we can work with unbounded contours as follows (see~\cite[Corollary~5.3]{stable}):
\begin{Prp} For any integer~$p \geq 1$, the solution of the Cauchy problem for the Teukolsky equation
with initial data~$\Psi|_{t=0} = \Psi_0 \in \D(H)$ has the representation
\beq \label{propagatorp}
\Psi(t) = -\frac{1}{2 \pi i} \int_C e^{-i \omega t}\: \frac{1}{(\omega + 3 i c)^p} \;
\Big( R_\omega \,\big(H + 3 i c \big)^p \,\Psi_0  \Big)\: d\omega \:,
\eeq
where~$C$ is the contour
\beq \label{Cdef}
C = \big\{ \omega \:\big|\: \im \omega = 2c \big\} \cup 
\big\{ \omega \:\big|\: \im \omega = - 2c \big\}
\eeq
with counter-clockwise orientation.
\end{Prp} \noindent
Here the factor~$(\omega + 3 i c)^{-p}$ gives suitable decay for large~$|\omega|$
and ensures that the integral converges in the Hilbert space~$\H$.

The representation~\eqref{propagatorp} gives an explicit solution of the Cauchy problem
in terms of a Cauchy integral of the resolvent. Unfortunately, this representation
does not immediately give information on the
long-time dynamics of the Teukolsky wave. This shortcoming can be understood immediately
from the fact that the factor~$e^{-i \omega t}$ in the integrand increases exponentially
for large times because~$|e^{-i \omega t}| = e^{\im \omega t} = e^{\pm 2 c t}$.
In order to bypass this shortcoming, our strategy is to move the contour onto the real axis.
Once this has been accomplished, the integral representation~\eqref{propagatorp} simplifies to a Fourier transform,
\[ \Psi(t) = \int_{-\infty}^\infty e^{-i \omega t}\: \hat{\Psi}(\omega) \:d\omega \:. \]
The decay of such a Fourier transform can be obtained from the {\em{Riemann-Lebesgue lemma}},
stating that
\[ \hat{\Psi} \in L^1(\R, d\omega) \quad \Longrightarrow \quad \lim_{t \rightarrow \pm \infty}
\Psi(t) = 0 \]
(where the wave functions are evaluated pointwise in space).
One of the difficulties in making this strategy work is to prove that the contour can indeed be
moved onto the real axis. This makes it necessary to show that the Hamiltonian has no spectrum
away from the real axis. We did not succeed in proving this result using operator theoretic methods.
Instead, our method is to first make use of the separation of variables, making it possible
rule out the spectrum in the complex plane using Whiting's mode stability result~\cite{whiting}.

\section{A Spectral Decomposition of the Angular Teukolsky Operator} \label{seccomplete}
Following the strategy we just outlined, our
next task is to employ the separation of variables in the integrand of
the integral representation~\eqref{propagatorp}.
Regarding the angular equation~\eqref{coupled}
as an eigenvalue equation, we are led to considering
the angular operator~$\A_\omega$ in~\eqref{Aop} as an operator on the
Hilbert space
\[ \H_k := L^2(S^2) \cap \{ e^{-i k \varphi}\: \Theta(\vartheta) \:|\: \Theta : (0, \pi) \rightarrow \C \} \]
with dense domain~$\D(\A_\omega) = C^\infty(S^2) \cap \H_k$.
Unfortunately, the parameter~$\omega$ is not real but lies on the contour~\eqref{Cdef}.
As a consequence, the operator~$\A_\omega$ is not symmetric,
because its adjoint is given by
\[ \A_\omega^* = \A_{\overline{\omega}} \neq \A_\omega \:. \]
The operator~$\A_\omega$ is not even a normal operator, 
making it impossible to apply the spectral theorem in Hilbert spaces.
Indeed, $\A_\omega$ does not need to be diagonalizable, because
there might be Jordan chains.
On the other hand, in order to make use of the separation of variables, we must decompose
the initial data into angular modes. This can be achieved by decomposing the
angular operator into invariant subspaces of bounded dimension, as is made
precise in the following theorem (see~\cite[Theorem~1.1]{tspectral}):
\begin{Thm} \label{thmmain}
Let~$U \subset \C$ be the strip
\[ |{\mbox{\rm{Im}}}\, \omega| < 3c \:. \]
Then there is a positive integer~$N$ and a family of bounded
linear operators~$Q^\omega_n$ on~$\H_k$
defined for all~$n \in \N \cup \{0\}$ and $\omega \in U$
with the following properties:
\begin{itemize}[leftmargin=2.5em]
\item[\rm{(i)}] The image of the operator~$Q^\omega_0$ is an $N$-dimensional invariant
subspace of $\A_k$. 
\item[\rm{(ii)}] For every $n\geq1$, the image of the operator~$Q^\omega_n$ is an at most
two-dimensional invariant subspace of $\A_k$.
\item[\rm{(iii)}] The $Q^\omega_n$ are uniformly bounded in~$\Lin(\H_k)$,
i.e. for all $n \in \N \cup \{0\}$ and~$\omega \in U$,
\[ \|Q^\omega_n\| \leq c_2 \]
for a suitable constant $c_2=c_2(s,k,c)$ (here~$\| \cdot \|$ denotes the $\sup$-norm on~$\H_k$).
\item[\rm{(iv)}] The~$Q^\omega_n$ are idempotent and mutually orthogonal in the sense that
\[ Q^\omega_n\, Q^\omega_{n'} = \delta_{n,n'}\: Q^\omega_n \qquad \text{for all~$n,n' \in \N \cup \{0\}$}\:. \]
\item[\rm{(v)}] The $Q^\omega_n$ are complete in the sense that for every~$\omega \in U$,
\beq \label{strongcomplete}
\sum_{n=0}^\infty Q^\omega_n = \1
\eeq
with strong convergence of the series.
\end{itemize}
\end{Thm} \noindent

\section{Invariant Disk Estimates for the Complex Riccati Equation}
In order to locate the spectrum of~$\A_\omega$, we use detailed ODE estimates.
The operators~$Q^\omega_n$ are then obtained similar to~\eqref{Cauchy}
as Cauchy integrals,
\[ Q^\omega_n := -\frac{1}{2 \pi i} \ointctrclockwise_{\Gamma_n} s_\lambda\: d\lambda
\:,\qquad n \in \N_0 \:, \]
where the contour~$\Gamma_n$ encloses the corresponding spectral points,
and~$s_\lambda=(\A_\omega - \lambda)^{-1}$ is the resolvent of the angular operator.
What makes the analysis doable is the fact that~$\A_\omega$ is an ordinary
differential operator. Transforming the angular equation in~\eqref{coupled} into
Sturm-Liouville form
\begin{equation} \label{5ode}
\left( -\frac{d^2}{du^2} + V(u) \right) \phi = 0 \:,
\end{equation}
(where~$u=\vartheta$ and~$V \in C^\infty((0, \pi), \C)$ is a complex potential),
the resolvent~$s_\lambda$ can be represented as an integral operator whose kernel
is given explicitly in terms of suitable fundamental solutions~$\phiD_L$ and~$\phiD_R$,
\beq \label{sldef}
s_\lambda(u,u') = \frac{1}{w(\phiD_L, \phiD_R)} \times \left\{
\begin{aligned} \phiD_L(u)\: \phiD_R(u') &\quad&& \text{if~$u \leq u'$} \\
\phiD_L(u')\: \phiD_R(u) &&& \text{if~$u' < u$}\:,
\end{aligned}  \right.
\eeq
where~$w(\phiD_L, \phiD_R)$ denotes the Wronskian.

The main task is to find good approximations for the solutions of the
Sturm-Liouville equation~\eqref{5ode} with rigorous error bounds
which must be uniform in the parameters~$\omega$ and~$\lambda$.
These approximations are obtained by ``glueing together'' suitable WKB, Airy and parabolic cylinder functions.
The needed properties of these special functions are derived in~\cite{special}.
In order to obtain error estimates, we combine several methods:
\begin{itemize}[leftmargin=2.5em]
\item[(a)] Osculating circle estimates (see~\cite[Section~6]{tspectral})
\item[(b)] The $T$-method (see~\cite[Section~3.2]{tinvariant})
\item[(c)] The $\kappa$-method (see~\cite[Section~3.3]{tinvariant})
\eitem
The method~(a) is needed in order to separate the spectral points of~$\A_\omega$ 
(gap estimates).
The methods~(b) and~(c) are particular versions of {\em{invariant disk}} estimates
as derived for complex potentials in~\cite{invariant} (based on previous estimates for real potentials
in~\cite{angular} and~\cite{wdecay}).
These estimates are also needed for the analysis of the radial equation, see Section~\ref{secres} below.
We now explain the basic idea behind the invariant disk estimates.

Let~$\phi$ be a solution of the Sturm-Liouville equation~\eqref{5ode} with
a complex potential~$V$. Then the function~$y$ defined by
\[ y = \frac{\phi'}{\phi} \]
is a solution of the Riccati equation
\begin{equation} \label{riccati}
y'= V-y^2\:.
\end{equation}
Conversely, given a solution~$y$ of the Riccati equation,
a corresponding fundamental system for the Sturm-Liouville equation is obtained
by integration. With this in mind, it suffices to construct a particular approximate
solution~$\tilde{y}$ and to derive rigorous error estimates.
The invariant disk estimates are based on the observation that the Riccati
flow maps disks to disks (see~\cite[Sections~2 and~3]{invariant}).
In fact, denoting the center of the disk by~$m \in \C$ and its radius by~$R>0$, we get the
flow equations
\begin{align*}
R' &= -2 R \; {\mbox{\rm{Re}}}\, m \\
m' &= V - m^2 - R^2 \:.
\end{align*}
Clearly, this system of equations is as difficult to solve as the original Riccati equation~\eqref{riccati}.
But suppose that~$m$ is an approximate solution in the sense that
\begin{align*}
R' &= -2 R \; {\mbox{\rm{Re}}}\, m + \delta R \\
m' &= V - m^2 - R^2 \:+\: \delta m\:,
\end{align*}
with suitable error terms~$\delta m$ and~$\delta R$, then the Riccati flow
will remain inside the disk provided that its radius grows sufficiently fast, i.e.\
(see~\cite[Lemma~3.1]{invariant})
\[ \delta R \geq |\delta m|\:. \]
This is the starting point for the invariant disk method.
In order to reduce the number of free functions, it is useful to
solve the linear equations in the above system of ODEs by integration.
For more details we refer the reader to~\cite{invariant, tinvariant}.

\section{Separation of the Resolvent and Contour Deformations} \label{secres}
The next step is to use the spectral decomposition of the angular operator in Theorem~\ref{thmmain}
in the integral representation of the solution of the Cauchy problem.
More specifically, inserting~\eqref{strongcomplete} into~\eqref{propagatorp} gives
\beq \label{propagator2}
\Psi(t) = -\frac{1}{2 \pi i} \int_C \;\sum_{n=0}^\infty e^{-i \omega t}\: \frac{1}{(\omega + 3 i c)^p} \;
\Big( R_\omega \,Q^\omega_n \,\big(H + 3 i c \big)^p \,\Psi_0  \Big)\: d\omega \:.
\eeq
At this point, the operator product~$R_\omega Q^\omega_n$ can be expressed in
terms of solutions of the radial and angular ODEs~\eqref{coupled} which arise in
the separation of variables (see~\cite[Theorem~7.1]{stable}).
Namely, the operator~$Q^n_\omega$ maps onto an invariant subspace of~$\A_\omega$
of dimension at most~$N$, and it turns out that the operator product~$R_\omega \,Q^\omega_n$ leaves this
subspace invariant. Therefore, choosing a basis of this invariant subspace,
the PDE~$(H-\omega)R_\omega Q^n_\omega = Q^n_\omega$ can be rewritten as
a radial ODE involving matrices
of rank at most~$N$. The solution of this ODE can be expressed explicitly in terms
of the resolvent of the radial ODE. In order to compute this resolvent, it is useful to also transform the radial
ODE into Sturm-Liouville form~\eqref{5ode}.
To this end, we introduce the Regge-Wheeler coordinate~$u \in \R$ by
\[ \frac{du}{dr} = \frac{r^2+a^2}{\Delta} \:, \]
mapping the event horizon to $u=-\infty$. Then the radial ODE takes
again the form~\eqref{5ode}, but now with~$u$ defined on the whole real axis.
Thus the resolvent can be written as an integral operator with
kernel given in analogy to~\eqref{sldef} by
\[ s_\omega(u,v) = \frac{1}{w(\acute{\phi}, \grave{\phi})} \:\times\:
\left\{ \begin{array}{cl} \acute{\phi}(u)\, \grave{\phi}(v) & {\mbox{if~$v \geq u$}} \\[0.3em]
\grave{\phi}(u)\, \acute{\phi}(v) & {\mbox{if~$v < u$\:,}} \end{array} \right. \]
where~$\acute{\phi}$ and~$\grave{\phi}$ form a specific fundamental system
for the radial ODE.
The solutions~$\acute{\phi}$ and~$\grave{\phi}$ are constructed as
Jost solutions, using methods of one-dimensional scattering theory
(see~\cite{alfaro+regge} and~\cite[Section~6]{stable}, \cite[Section~3]{wdecay}).

The next step is to deform the contour in the integral representation~\eqref{propagator2}.
Standard arguments show that the integrand in~\eqref{propagator2} is holomorphic
on the resolvent set (i.e.\ for all~$\omega$ for which the resolvent~$R_\omega$ in~\eqref{propagatorp}
exists).
Thus the contour may be deformed as long as it does not cross singularities of the resolvent.
Therefore, it is crucial to show that the integrand in~\eqref{propagator2} is meromorphic
and to determine its pole structure.
Here we make essential use of Whiting's mode stability result~\cite{whiting}
which states, in our context, that every summand in~\eqref{propagator2} is holomorphic
off the real axis. In order to make use of this mode stability, we need to interchange the
integral in~\eqref{propagator2} with the infinite sum.
To this end, we derive estimates which show that the summands in~\eqref{propagator2}
decay for large~$n$ uniformly in~$\omega$. Here we again use
ODE techniques, in the same spirit as described above for the angular equation
(see~\cite[Section~10]{stable}).
In this way, we can move the contour in the lower half plane arbitrarily close to the real axis. Moreover,
the contour in the upper half plane may be moved to infinity. We thus obtain
the integral representation (see~\cite[Corollary~10.4]{stable})
\[ \Psi(t) = -\frac{1}{2 \pi i} \sum_{n=0}^\infty\:\lim_{\varepsilon \searrow 0}
\int_{\R - i \varepsilon} \frac{e^{-i \omega t}}{(\omega + 3 i c)^p} \;
\Big( R_{\omega,n}\:Q_n^\omega \,\big(H + 3 i c \big)^p \,\Psi_0  \Big)\: d\omega \:. \]

The remaining issue is that the integrands in this representation might have 
poles on the real axis.
These so-called {\em{radiant modes}} are ruled out by a causality argument 
(see~\cite[Section~11]{stable}; for an alternative proof see~\cite{andersson+whiting}). We thus obtain the following result (see~\cite[Theorem~12.1]{stable}).
\begin{Thm} \label{thmrep}
For any~$k \in \Z/2$, there is a parameter~$p>0$ such that for any~$t<0$, the solution of the Cauchy problem 
for the Teukolsky equation with initial data
\[ \Psi|_{t=0} = e^{-i k \varphi}\: \Psi_0^{(k)}(r, \vartheta) \qquad \text{with} \qquad
\Psi^{(k)}_0 \in C^\infty(\R \times S^2, \C^2) \]
has the integral representation
\beq \begin{split}
\Psi&(t,u,\vartheta, \varphi) \\
&= -\frac{1}{2 \pi i} \:e^{-i k \varphi}\: \sum_{n=0}^\infty \int_{-\infty}^\infty \frac{e^{-i \omega t}}{(\omega + 3 i c)^p} \;
\Big( R^-_{\omega,n} \:Q_n^\omega
\big(H + 3 i c \big)^p \,\Psi_0^{(k)}  \Big)(u, \vartheta)\: d\omega \:,
\end{split} \label{propfinal}
\eeq
where~$R^-_{\omega,n} \Psi := \lim_{\varepsilon \searrow 0} \big(R_{\omega-i \varepsilon,n} \Psi)$.
Moreover, the integrals in~\eqref{propfinal} all exist in the Lebesgue sense.
Furthermore, for every~$\varepsilon>0$ and~$u_\infty \in \R$, there is~$N$ such that
for all~$u<u_\infty$,
\beq \label{biges}
\sum_{n=N}^\infty \int_{-\infty}^\infty \bigg\| \frac{1}{(\omega + 3 i c)^p} \;
\Big( R^-_{\omega,n} \:Q_n^\omega
\,\big( H + 3 i c )^p \,\Psi_0^{(k)} \Big)(u) \bigg\|_{L^2(S^2)} \: d\omega < \varepsilon \:.
\eeq
\end{Thm}

\section{Proof of Pointwise Decay} \label{secradial}
Theorem~\ref{thmdecay} is a direct consequence of the integral representation~\eqref{propfinal}
in Theorem~\ref{thmrep}. Namely, combining the estimate~\eqref{biges} with Sobolev methods,
one can make the contributions for large~$n$ pointwise arbitrarily small.
On the other hand, for each of the angular modes~$n=0,\ldots, N-1$, 
the desired pointwise decay as~$t \rightarrow -\infty$ follows from the Riemann-Lebesgue lemma.
For details we refer to~\cite[Section~12]{stable}.

\section{Concluding Remarks} \label{secoutlook}
We first point out that the integral representation of Theorem~\ref{thmrep} is
a suitable starting point for a detailed analysis for the dynamics of the solutions of the Teukolsky equation.
In particular, one can study decay rates (similar as worked out for massive Dirac waves in~\cite{wdecay})
and derive uniform energy estimates outside the ergosphere (similar as for scalar waves in~\cite{sobolev}).
Moreover, using the methods in~\cite{penrose}, one could analyze superradiance phenomena
for wave packets in the time-dependent setting.

Clearly, the next challenge is to prove {\em{nonlinear stability}} of the Kerr geometry.
This will make it necessary to refine our results on the linear problem, for example
by deriving weighted Sobolev estimates and by analyzing the $k$-dependence of our estimates.
Moreover, it might be useful to combine our methods and results with microlocal techniques
(as used for example in the proof of nonlinear stability results in the related
Kerr-De Sitter geometry~\cite{hintz+vasy}).

\Thanks {{\em{Acknowledgments:}}
I would like to thank the organizers of the first
{\sf{``Domoschool -- International Alpine School of Mathematics and Physics''}}
held in Domodossola, 16-20 July 2018, for the kind invitation.
This article is based on my lectures delivered at this summer school.
I am grateful to Niky Kamran and Igor Khavkine for helpful comments on the manuscript.

\providecommand{\bysame}{\leavevmode\hbox to3em{\hrulefill}\thinspace}
\providecommand{\MR}{\relax\ifhmode\unskip\space\fi MR }
\providecommand{\MRhref}[2]{%
  \href{http://www.ams.org/mathscinet-getitem?mr=#1}{#2}
}
\providecommand{\href}[2]{#2}

\end{document}